\documentclass[a4paper,twocolumn]{revtex4}
\usepackage{amsmath}
\usepackage{amsfonts}
\usepackage{amssymb}
\usepackage{graphicx}
\usepackage{subscript}
\usepackage{siunitx}
\usepackage{xcolor}
\usepackage{gensymb}
  
\begin{document}
\title{Design of patchy rhombi: from close-packed tilings to open lattices}

\author{Carina Karner}
\email{carina.karner@univie.ac.at}
\affiliation{Faculty of Physics, University of Vienna, Boltzmanngasse 5, A-1090, Vienna, Austria}

\author{Christoph Dellago}
\affiliation{Faculty of Physics, University of Vienna, Boltzmanngasse 5, A-1090, Vienna, Austria}

\author{Emanuela Bianchi}
\email{emanuela.bianchi@tuwien.ac.at}
\affiliation{Institut f{\"u}r Theoretische Physik, TU Wien, Wiedner Hauptstra{\ss}e 8-10, A-1040 Wien, Austria }
\affiliation{CNR-ISC, Uos Sapienza, Piazzale A. Moro 2, 00185 Roma, Italy}

\date{\today}

\begin{abstract} 
In the realm of functional materials, the production of two-dimensional structures with tunable porosity is of paramount relevance for many practical applications: surfaces with regular arrays of pores can be used for selective adsorption or immobilization of guest units that are complementary in shape and/or size to the pores, thus achieving, for instance, selective filtering or well-defined responses to external stimuli. The principles that govern the formation of such structures are valid at both the molecular and the colloidal scale. Here we provide simple design directions to combine the anisotropic shape of the building units -- either molecules or colloids -- and selective directional bonding. Using extensive computer simulations we show that regular rhombic platelets decorated with attractive and repulsive interaction sites lead to specific tilings, going smoothly from close-packed arrangements to open lattices. The rationale behind the rich tiling scenario observed can be described in terms of steric incompatibilities, unsatisfied bonding geometries and interplays between local and long-range order.
\end{abstract}

\maketitle

\section{Introduction}

The interest in self-assembled monolayers stems from the variety of technological applications they can be designed for. At the molecular scale~\cite{COCIS_2009,NatMat_2009,RSCadv_2013,NatComm_2017},  properties such as corrosion resistance, surface superhydrophobicity, or antifouling are imparted to surfaces via close-packed tilings, while porous surface networks are at the basis of the so called two-dimensional host-guest chemistry~\cite{COCIS_2009,NatMat_2009,RSCadv_2013,NatComm_2017}.  At the colloidal scale, while close-packed tilings are mostly used as lithographic masks~\cite{SoftMatter_2012,Nanotoday_2018} or as seeds for the so-called colloidal epitaxy~\cite{Nature_97}, open lattices offer many possibilities for the rational design of selective membranes~\cite{Nanolett_2011,ChemComm_2015} or two-dimensional materials with optical, magnetic, electronic or catalytic properties~\cite{Nanoscale_2014,CSR_2014,CR_2016}.

The  formation  of  complex surface  patterns at the molecular scale relies on the hierarchical assembly of organic functional molecules to self assembled monolayers on solid substrates, such as graphite or metallic surfaces~\cite{NatChem_2010,NatChem_2012}. 
Such assemblies of molecular species that lie flat on relatively inert substrates mostly result from reversible non-covalent interactions: self-similar, fractal aggregates, nonporous  -- ordered as well as disordered -- networks and even quasi-crystalline patterns emerge due to hydrogen~\cite{Nature_2003,Science_2008,NatChem_2012,JACS_2013,Nature_2014} or halogen~\cite{JPCC_2017,NatComm_2018} bonding or van der Waals interactions~\cite{ACSNano_2012,JACS_2013_bis}. 
Beyond the atomic details, a combination of simple factors fully describes the self-assembly of many different surface patterns: the geometric features of the building units, such as the aspect ratio or the rotational symmetry of the molecules, together with the placement of the bonding groups and possible energy differences between binding groups mostly determine the features of the assembled monolayers~\cite{Whitelam2012,Whitelam2015,CollSurfA_2017}. For instance, the assembly of elongated organic molecules functionalized with four carboxyl groups on a graphene surface can be described by rhombic platelets with four interactions sites~\cite{Whitelam2012}: molecules of this type bind to each other in two possible orientations (parallel and non-parallel) and the steric incompatibilities between rombic platelets -- combined with the presence of bonding sites in suitable arrangements -- are able to reproduce the directionality of the bonds between the carboxyl groups (see panels (A) to (F) in Fig.~\ref{fig:feq_phases}). 

The same principles are applicable at the colloidal level. Nowadays, monodisperse non-spherical particles at the nano- and micro-scale are largely available~\cite{Sacanna2011,Ravaine2013,Vutukuri2014,Miriam2015,Small_2018}, thus allowing to create a vast variety of structures with different symmetries and packing densities~\cite{DeGraaf2012,Damasceno2012,Schultz2015}. While two-dimensional hard shapes form entropically stabilized lattices that tend to maximize edge-to-edge contact~\cite{Ye2013,Millan2014,Lee_2016}, the presence of surface binding groups introduces a competition between entalphy and entropy that can stabilize open architectures~\cite{Millan2014}.

While the self-assembly of surface structures can be studied by synthesizing many candidate building blocks with different geometries and interactions in the laboratory, computer simulations of simple models offer an efficient way to 
reveal the basic principles governing the assembly process.

The great advantage of such a numerical approach is that it allows to quickly explore a huge parameter space without the need to manufacture the -- possibly unsuccessful -- building units and for this reason computer simulations have been used extensively in materials design~\cite{Whitelam2012,Ye2013,Millan2014,CollSurfA_2017}. 

Here, we propose a class of anisotropic planar units provided with a fixed and low number of interaction centers and we investigate how the tiling ability of these units depends on the nature and the arrangement of the interaction centers using Monte Carlo (MC) Simulations.  Within this framework, specific porous structures -- with target pores -- can be stabilized with respect to others, just by taking advantage of few selected characteristics. 
While porous monolayers with a specific pore size have been achieved -- experimentally as well as in simulations -- by a careful blending of selected features (such as the anisotropy of the particle shape and/or bonding patterns)~\cite{Granick_2011,Whitelam2012,Ye2013,Millan2014} our approach shows how to switch continuously from an open to a close-packed tiling.

\section{Methods}

\subsection{Particle model}
In this work we explore the assembly products of regular hard rhombi in two dimensions with four square-well interaction sites (referred to as patches) of either one type denoted as four-equal (feq) or of two types denoted as double-manta, double-mouse and checkers (dma/dmo/checkers). Patches of the same kind attract each other with strength $-\epsilon$ and patches of a different kind repel each other with $\epsilon$.

The interaction potential between two hard rhombi $i$ and $j$ is $0$ whenever they do not overlap and infinity if they do overlap: 
\[ U(\vec{r}_{ij}, \Omega_{i}, \Omega_{j})  =
  \begin{cases}
    0     & \quad \text{if  $i$ and $j$ do not overlap}\\
    \infty  & \quad \text{if $i$ and $j$ do overlap}.\\
  \end{cases}
\]
where $\vec{r}_{ij}$ is the distance vector, and $\Omega_{i}$ and $\Omega_{j}$ denote the particle orientations. The patches exhibit an attractive or repulsive square-well type potential:
\[ W(p_{ij})  =
  \begin{cases}
    \pm \epsilon     & \quad \text{if}\quad p_{ij}< 2r_{p}\\
    0 & \quad  \text{if}\quad p_{ij} \geq 2r_{p}, \\
  \end{cases}
\]
where $p_{ij}$ is the patch-patch distance vector, $2r_{p}$ is the patch diameter and $\epsilon$ denotes the patch interaction strength.
Note that within a Monte-Carlo (MC) routine a hard particle orientation/translation move is accepted if particles do not overlap. Hence, the calculation of the hard particle potential requires a collision detection check between pairs of particles using the separating axis theorem~\cite{Golshtein_1996} (for convex particles only).
In summary, the MC-evaluation of the total energy consists of a collision detection check and the subsequent calculation of the patch pair potential.
 
For every class of patch specificities (feq/dma/dmo/checkers) we explore different patch topologies, which enables us to explore the patch parameter space systematically (see Appendix ~\ref{patch-topologies} for details on the different topologies).
For all studied topologies we are able to describe the specific 
patch positioning with one variable, $\Delta$.
Table \ref{table:geom} gives a full overview of the particle parameters.

\begin{table}[h]
\begin{center}
\begin{tabular}{ |l|l|l| } 
\hline
 \bf{parameter} & \bf{symbol} & \bf{value} \\
 \hline
 angle & $\alpha$ & $60\degree$ \\ 
 \hline
 side length & l & 1.0 \\ 
 \hline
 patch radius & $r_{p}$  & 0.05 \\
 \hline
 interaction strength & $\epsilon$ & $\pm [4.2, 4.8, 5.2]$ $k_{B}T$\\
 \hline
 patch position  & $\Delta$ & [0.2,0.3,0.4,0.5,0.6,0.7,0.8] \\
 \hline
\end{tabular}
\caption{Particle parameters. Note that patch position refers to the relative placement of a patch on a rhombi edge. See Fig.~\ref{fig:feq_phases} and Fig.~\ref{fig:phases}.}\label{table:geom}
\end{center}
\end{table}

\subsection{Simulation details}
To model the absorption of platelets on a surface we choose the grand canonical ensemble ($\mu VT$) with single particle rotation and translation moves and particle insertion and deletion. To avoid kinetic traps we implement cluster moves \cite{Whitelam2007, Whitelam2010} (see Appendix~\ref{cluster-move-algorithm} for details).
The shape of the rectangular box is chosen such that a close-packed tiling of rhombi fits fully. The used system parameters are summarized in  table \ref{table:system_param}.
We follow the same procedure in all investigated systems: after equilibration of $3\times10^5$ MC-sweeps in a regime of very low packing fractions of $\phi\approx 0.05$ with $\mu_{eq}$, we increase the chemical potential to $\mu^{*}$ to observe the assembly. In total we perform 8 simulations per state point ($\epsilon$, $\Delta$). 
The number of MC-sweeps until equilibration varies from system to system and lies between $\approx 1.2\times 10^6 - 2.0\times 10^7$ MC-sweeps (see Fig.\ref{fig:feq_npt} in Appendix~\ref{equilibration}).
Typical system sizes of the completed assemblies are in the range of $N\approx 1000$ with $N$ as the number of particles, while the area is fixed $A=1000\cdot\sin{(60\degree)}$. Numerical values of the system parameters are given in the supporting information. 
We characterize the emerging tilings with the randomness parameter $\langle \Psi \rangle$ of the largest cluster, averaged over all simulation runs (see panel (G) of Fig.~\ref{fig:feq_phases}, panel (B) of Fig.~\ref{fig:domain_sizes} and Fig.~\ref{fig:domain_size_histo} in Appendix~\ref{domain-sizes}).

\begin{table}[h]
    \begin{center}
    \begin{tabular}{|l|l|l|}
        \hline
        \bf{system parameter} &  \bf{symbol} & \bf{value} \\
        \hline
         Area & A & $1000 \cdot \sin{(60\degree)}$ \\
         \hline
         box width & $L_{x} $ & $\sqrt{1000}$ \\
         \hline
         box height & $L_{y}$ &  $\sqrt{1000}
         \cdot\sin{(60\degree)}$ \\
         \hline
         chemical potential eq. & $\mu_{eq}$ & 0.1 \\
         \hline
         chemical potential & $\mu^{*}$ & 0.25 \\
         \hline
         Boltzmann constant  & $k_{\rm B}$ & 1 \\
         \hline
         Temperature & $T$ & 0.1\\ 
         \hline
          fictitious inverse $T$  & $\beta_{f}$ & 5 \\ 
         \hline
    \end{tabular}
    \caption{The system parameters used in all simulations.}
    \label{table:system_param}
    \end{center}
\end{table}


\section{Results}

\begin{figure*}
\begin{center}
\includegraphics[width=\textwidth]{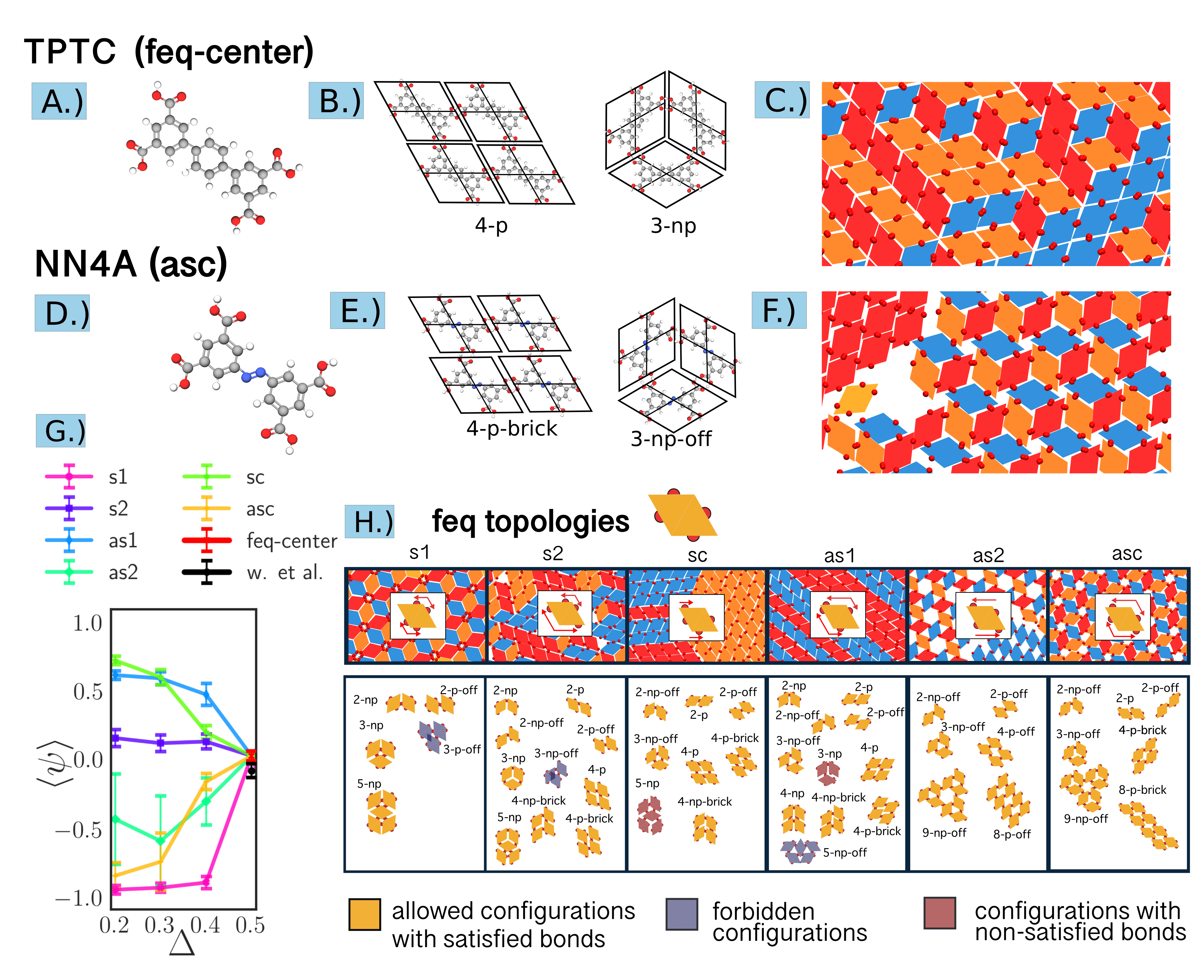}
\caption{First row: (A) molecular structure of TPTC (p-terphenyl-3,5,3',5'- tetracarboxylic acid), (B) two possible arrangements of TPTC molecules linked via hydrogen bonds: four molecules in a parallel (4-p) arrangement (left) and three molecules in a non-parallel (3-np) arrangement (right), (C) simulation snapshot of the tiling formed by regular rhombi with four equal patches placed at $\Delta=0.5$ (feq-center); particle colors in the simulation snapshot refer to the particle orientation. Second row: (D) molecular structure of 3,3,3',5'-azobenzene tetracarboxylic acid  (NN4A),
(E) two possible arrangements of NN4A molecules linked via hydrogen bonds: four molecules in a parallel (4-p-brick) arrangement  (left)  and three molecules in a non-parallel (3-np-off) arrangement (right), (F) simulation snapshot of the tiling formed by  feq particles with patches arranged in the so-called asc topology with $\Delta=0.3$. Third row, panel (G), average order parameter $\langle\Psi\rangle$ as function of $\Delta$, where -1 denotes a completely non-parallel tiling, 1 a completely parallel tiling and 0 a random tiling; six different patch topologies are considered, namely s1, s2, sc, as1, as2, and asc, as labeled. Third row, panel (H), upper part: simulations snapshots for all feq systems; in the snapshots all different patch topologies have $\Delta=0.2$.  The color of the particles in the snapshots highlights their orientation. The inset in each panel highlights the patch topology by means of arrows whose length is proportional to the distance between the patch and its vertex. We note that the feq-center particle is reported above the snapshots as reference. Third row, panel (h), lower part: for each patch topology, we consider small clusters of particles that tile with all bonds satisfied (yellow), that cannot tile because of overlaps (gray) and that tile with unsatisfied bonds (burgundy). Configurations are labeled according to the number of particles in the cluster (from 2 up to 9), their bond arrangement -- parallel (p) or non-parallel (np) -- and  their respective positions -- edge-to-edge (no label), off-edge (off) and brick-like (brick). We note that the chemical potential is the same for each system and the final density of each sample is system-dependent (see Fig.~\ref{fig:feq_npt} in Appendix~\ref{equilibration}).}
\label{fig:feq_phases}
\end{center} 
\end{figure*}

\subsection{Feq systems}
The starting point of our explorations is the patchy rhombi system introduced by Whitelam {\it et al.}~\cite{Whitelam2012} to describe the assembly of small organic TPTC molecules ((p-terphenyl-3,5,3',5'- tetracarboxylic acid, see panel (A) of Fig.~\ref{fig:feq_phases}) on graphite. In general, tectons or building blocks of tetracarboxylic acids adsorb flat (with their aromatic backbone parallel to the substrate) on highly oriented pyrolytic graphite due to $\pi$ stacking interactions~\cite{JACS_2007,ChemCom_2008,Angew_2008,Science_2008,NatChem_2012}.

In Ref.~\cite{Whitelam2012}, TPTC molecules are modeled as hard regular rhombi (with internal angles $60\degree$ and $120\degree$) decorated with four patches that mimic the hydrogen bonding groups: each patch is placed at the center of an edge, adjacent rhombi can bond together only via their patches and all pairs of bonded patches bear the same energy. Despite the simplicity of the model referred to as feq-center (four equal patches at the edge centers), the assembly of these units on a featureless two-dimensional substrate successfully reproduced~\cite{Whitelam2012} the assembly of TPTC molecules into a random rhombus tiling~\cite{Science_2008}. The randomness  of a tiling is measured by a suitably designed order parameter $\Psi$~\cite{NatChem_2012}., that goes from +1 (p) to -1 (np), where  $\Psi=0$ is the perfect random tiling (see Methods for the definition of $\Psi$).
The feq-center system is our starting and reference point: it forms a nearly perfect random tiling with $\Psi\approx 0$. We note that, while Whitelam $et$ $al.$ observed a slightly non-parallel tiling with $\langle\Psi_{\text{w. et al.}}\rangle = -0.08 \pm 0.05$, in our simulations we obtain a slightly parallel tiling, $i.e.,$ $\langle\Psi_{\text{feq-center}}\rangle= 0.02\pm0.04$ (see panels (C) and (G) of Fig.~\ref{fig:feq_phases}). These two values are comparable within their error bars. 

While the effect of perturbing the aspect ratio of the rhombi has been investigated in Ref.~\cite{Whitelam2012}, little is known on the effect of the patch topology on the tiling motifs. At the molecular scale, a change in the patch topology can be achieved by considering different tectons of tetracarboxylic acids. In order to describe 3,3,3',5'-azobenzene tetracarboxylic acid  (NN4A) molecules~\cite{Angew_2008},  for instance, we can consider regular rhombi with two of the patches placed at the center of opposite edges and the other two patches positioned symmetrically out of center in opposite direction with respect to each other (see panel (D) of Fig.~\ref{fig:feq_phases}); we refer to units with this patch topology as feq-asc (four equal patches with an asymmetric and a centered pair of opposite patches). The same approach can be used for other tectons of tetracarboxylic acids on graphite~\cite{JACS_2007,ChemCom_2008,Angew_2008,Science_2008,NatChem_2012}, resulting in different patch arrangements. 

Within our model, regular rhombi can be arbitrarily decorated with four equal patches along the edges. To systematically explore this -- in principle infinite  -- parameter space, we describe each particle type by one single parameter $\Delta$ and a patch topology (see Methods and Fig.~\ref{fig:feq_phases} for the definition of the different topologies). 

In panel (H) of Fig.~\ref{fig:feq_phases} (upper part) we show the resulting tilings for all six patch topologies with $\Delta=0.2$, corresponding to the most extreme off-center cases (for a larger view on the systems see Fig.~\ref{fig:feq_npt} in Appendix~\ref{equilibration}). The visual analysis of the simulation snapshots suggests that four of six tilings yield close-packed lattices, while two lead to open tilings.

\subsubsection{Closed-packed feq tilings}

The four closed-packed tilings are characterized by very different degrees of randomness: at $\Delta=0.2$, $\langle\Psi_{\text{feq-s1}}\rangle=  -0.854 \pm 0.095$, $\langle\Psi_{\text{feq-s2}}\rangle= 0.155 \pm 0.064.$,
$\langle\Psi_{\text{feq-sc}}\rangle= 0.725 \pm 0.032$ and  $\langle\Psi_{\text{feq-as1}}\rangle= 0.617 \pm 0.030$
(see panel (G) of Fig.~\ref{fig:feq_phases}). The differences between these tilings can be understood by considering
small clusters of particles, reported in panel (H) of Fig.~\ref{fig:feq_phases} (lower part).
Essentially, the degree of randomness of a tiling rises with the compatibility of these small clusters with p-
and np-bonding arrangements.
While at the two-particle level all four patch topologies can bond parallel (p) as well as non-parallel (np), already at the three- and then five-particle level, bonding configurations with either unsatisfied bonds or particle overlaps emerge.

In the feq-s1 system the small cluster analysis shows that p-clusters larger than two are not possible due to overlaps, leading to a mostly np-tiling. We note that, even though the long-range order is purely non-parallel  $\langle\Psi_{\text{feq-s1}}\rangle \neq -1$ due to the presence of grain boundaries between different np-domains. In contrast, in the feq-s2 both p- and np-bonding arrangements are compatible with local as well as long-range order, resulting in a random tiling. It is worth noting that, in this case, too many tiling possibilities make a hole-free tiling highly unlikely.
Finally, feq-sc and feq-as1 are characterized by the presence of large p-domains connected to each other within a roof-shingle motif: in the first case clusters with unsatisfied bonds emerge at the level of five-particle clusters, while in the second case clusters with unsatisfied bonds emerge already at the level of three-particle clusters. When $\Delta$ is increased from 0.2 towards 0.5 (see panel (G) of Fig.~\ref{fig:feq_phases}), all random tilings monotonically collapse into the nearly perfect random tiling corresponding to the feq-center one. In general as $\Delta$ grows towards 0.5 a closer analysis of the configurations shows a growing commensurability of the various local bonding patterns. Moreover, the number of defects -- $i.e.,$ the number of holes in a close-packed tiling -- reduces as $\Delta$ grows towards 0.5 (see panel (G) of Fig.~\ref{fig:feq_phases}).

\subsubsection{Open feq tilings}

A different scenario emerges for feq-as2 and feq-asc, for which open lattices are formed. The differences between these tilings can be understood by considering small clusters of particles, reported in panel (H) of Fig.~\ref{fig:feq_phases} (lower part). As the small cluster analysis highlights (see panel (H) of Fig.~\ref{fig:feq_phases}), both patch topologies allow the formation of p- and np-clusters with holes that are able to tile and ultimately yield open tilings. 

The emerging parallel tilings are characterized by either an open pattern with rhombic pores -- denoted as op-tiling -- (in the case of feq-as2) or by a closed-packed brick pattern (in the case of feq-asc); in contrast both non-parallel tilings are open lattices characterized by hexagonal and triangular pores -- denoted as onp-tilings. Note that the pore size of both tilings is dependent on $\Delta$, with pores largest at the extremes, $i.e.$, at $\Delta=0$ and $\Delta=1$, while at $\Delta=0.5$ the pores close and the systems collapse to the random tiling of feq-center. 

In feq-as2, open p- and np-clusters grow next to each other within the same sample leading to $\langle\Psi_{\text{feq-as2}}\rangle=  -0.439 \pm 0.332$ for $\Delta=0.2$. However, it is important to note that for such an extreme $\Delta$-value p- and np-domains are highly incompatible and avoid contact by paying a high energy price, hence we cannot classify this as a random tiling (see snapshot in FIG.~\ref{fig:feq_phases}, panel (H)). On increasing $\Delta$, a non-monotonous behavior is observed (see panel (G) of Fig.~\ref{fig:feq_phases}):  first $\langle\Psi_{\text{feq-as2}}\rangle$ drops down and then rises again to reach the value of the nearly perfect random tiling at $\Delta=0.5$. The sweet spot for the formation of the onp-tiling is observed at $\Delta=0.3$, where $\langle\Psi_{\text{feq-as2}}\rangle=  -0.600 \pm 0.332$.  At this $\Delta$-value, three of 16 runs show a pure onp-tiling, while the other 13 runs retain the coexistence between p- and np-domains (see the corresponding simulation snapshot in panel (H) of Fig.~\ref{fig:feq_phases} and Fig.~\ref{fig:feq_npt} in Appendix~\ref{equilibration}).

In contrast, the sweet spot for the formation of the onp-tiling for feq-asc is observed at $\Delta=0.2$, where $\langle\Psi_{\text{feq-asc}}\rangle= -0.854 \pm 0.095$. In this case, on increasing $\Delta$, the trend towards the nearly perfect random tiling at $\Delta=0.5$ is monotonous (see panel (G) of Fig.~\ref{fig:feq_phases}).

In both systems p- and np-domains become more commensurable as $\Delta$ increases. While at $\Delta = 0.2$, the energy price related to the coexistence of p- and np-domains is high, at $\Delta$=0.4 we already observe fully bonded domains with mixed orientation. It is worth noting that the energy price for such a coexistence is not only related to the patch arrangement but also to the bonding energy (see Appendix~\ref{energy-impact} for details). 

Within our description, NN4A molecules can be depicted as feq-asc particles with $\Delta = 0.3$: for such a system, even though both p- (4-p-brick) and np- (3-np-off) arrangements are allowed  (see panel (E) of Fig.~\ref{fig:feq_phases}), we observe the formation of an open tiling with a very strong onp-component (in the snapshot reported in panel (F) of Fig.~\ref{fig:feq_phases}, $\Psi_{\text{feq-asc}}=  -0.590946$). This result is consistent with experimental data~\cite{Angew_2008}.

\begin{figure*}
\begin{center} 
\includegraphics[width=\textwidth]{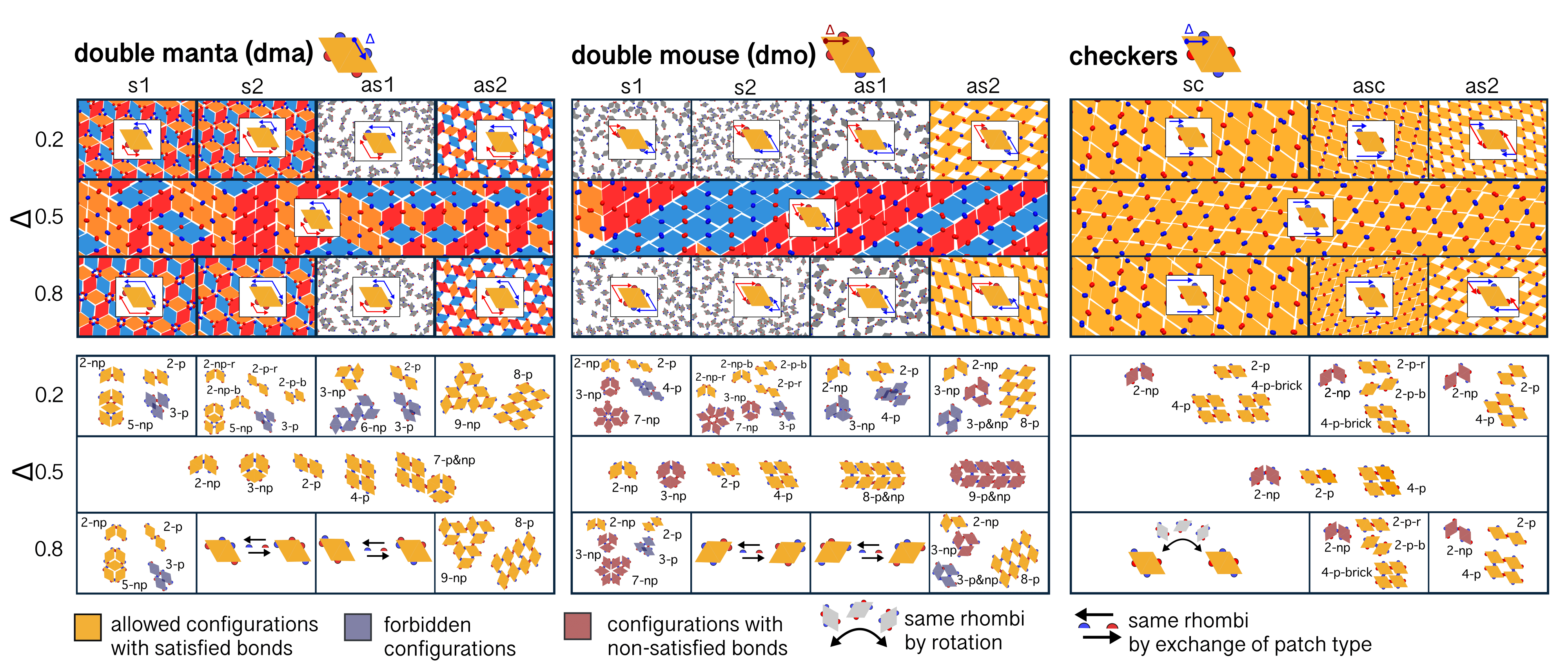}
\caption{Upper row: Simulation snapshots of resulting tilings for double manta (dma), double mouse (dmo) and checkers. The reference patch topologies for these classes of rhombi -- namely, dma-center, dmo-center and checkers-center -- are reported above the snapshots. For each class of particles, the columns show the different patch topologies (s1, s2, sc, as1, as2, and asc, each represented in the corresponding inset on the tiling snapshots) and the rows the different patch positions, determined by $\Delta$ (as labeled). The color of the particles in the snapshots highlights their orientation. Lower row: for each particle type, we consider small clusters with all bonds satisfied (yellow), with overlaps (gray) and with unsatisfied bonds (burgundy). When particles are the same by rotation or exchange of particles (see legend) small clusters are not shown. Configurations are labeled according to the number of particles in the cluster (from 2 up to 9), their bond arrangement -- parallel (p), non-parallel (np) or mixed (p$\&$np)-- and their respective positions -- edge-to-edge (no label) or brick-like (brick). In few cases we use -r and -b to distinguish two-particle clusters with bonds between either red or blue patches. We note that the chemical potential is the same for each system and the final density of each sample is system-dependent.}
\label{fig:phases}
\end{center} 
\end{figure*}

\subsection{Dma/Dmo/Checkers systems}
Within our minimal design approach, we propose to take into account the role of the patch identity. Tricarboxylic acid molecules on a gold surface were shown, for instance, to assemble into different structures according to their deprotonation level~\cite{ChemComm_2018}. By playing with the pH, $e.g.$, two types of interaction centers might be induced, whose relative abundance is determined by the deprotonation process. 

We thus consider now patchy rhombi with two kinds of patches for any patch arrangement (given again by a patch topology and a $\Delta$-value, see Fig.~\ref{fig:phases} and Methods for a detailed description). To describe how different tilings arise from the underlying patch topologies, we refer to Fig.~\ref{fig:phases} and divide results into three groups: center, symmetric off-center and asymmetric off-center patch topologies. 

In general, we note that the off-center tilings are symmetric with respect to $\Delta = 0.5$. Within a specific off-center topology,  the tiling geometry is -- most of the time -- preserved over the whole $\Delta$-range. However, in most cases, the particles yielding those tilings are not the same, but some are equivalent to each other (see Fig.~\ref{fig:phases} and the Methods section).

\subsubsection{Tilings of center topologies.}
We consider systems where patches are placed in the edge center ($\Delta=0.5$), referred to as dma-center, dmo-center and checkers-center. All particle types yield close-packed, space-filling tilings, but each patch topology induces specific tiling features (see the corresponding simulation snapshots in Fig.~\ref{fig:phases}, upper part) characterized by very different values of $\langle\Psi\rangle$.  

Similar to feq-center, dma-center yields an almost perfect random tiling with $\langle\Psi_{\rm dma-center} \rangle= 0.03 \pm 0.05$ (see panel (B) of Fig.~\ref{fig:domain_sizes}). As the small cluster analysis highlights, both p- and np-bonding arrangements are compatible with long-range order: the double manta specificity allows for multi-particle p-, np- as well as mixed clusters at any cluster size (see the corresponding panel in Fig.\ref{fig:phases}, lower part).

A different degree of randomness is observed in dmo-center, where p-bonds dominate leading to $\langle\Psi_{\rm dmo-center}\rangle=0.52\pm 0.02$ (see panel (B) of Fig.~\ref{fig:domain_sizes}). In this case, although both p- and np- pair bonds are allowed, closed loops of np-bonds (in the following referred to as boxes~\cite{Whitelam2012}) induce an energetic penalty (see, $e.g.$, the 3-np configuration in the corresponding panel of Fig.~\ref{fig:phases}, lower part). This restricts the phase to either p-clusters (see, $e.g.$, the 4-p configuration), roof-shingles (see, $e.g.$, the 8-p$\&$np configuration) or mixtures of those. Note however, that the close-off on roof-shingle motives inevitably induces either energetic penalties (see, $e.g.$, the 9-p$\&$np configuration) or defects (see the simulation snapshots in the corresponding panel of Fig.~\ref{fig:phases}, upper part, whenever three orientations coalesce). 

We note that the spatial distribution of p- and np-bonds -- and the degree of randomness of the resulting tiling -- can be quantified also by the average size of p- or np-domains, $\langle \sigma_p \rangle $ and $\langle \sigma_{np} \rangle$, together with the fraction np-domains, $\langle f^d_{np} \rangle $ (see panel (C) in Fig.~\ref{fig:domain_sizes} and Fig.~\ref{fig:domain_size_histo} in Appendix~\ref{domain-sizes}). 

A completely different lattice is observed for checkers-center.
As patches of the same kind sit on opposing edges, only p-bonds are possible without energetic penalties for all $\Delta$ values (see configuration 2-p $versus$ 2-np in the corresponding panel of Fig.~2, lower part).
Checkers-center yields a defect-free close-packed p-tiling with $\langle\Psi_{\rm checkers-center}\rangle = 1.00$ (see panel (B) of Fig.~\ref{fig:domain_sizes}).

\subsubsection{Tilings of symmetric off-center topologies.}

Dma and dmo symmetric off-center topologies result in very different tilings compared to their respective center topologies: 
placing the patches symmetrically off-center in those systems leads to p-bonds with offset edge contacts. Such a bonding arrangement forbids the growth of  multi-particle p-clusters in both classes of systems (see the small cluster analysis in the corresponding panels of Fig.~\ref{fig:phases}, lower part, where particle overlaps are highlighted). As in dma-s1/s2 the formation of boxes is still possible (see, $e.g.$, the 5-np configuration in the corresponding panels of Fig.~\ref{fig:phases}, lower part), defect-free np-tilings are observed with $\langle\Psi_{\rm dma-s1}\rangle=\langle\Psi_{\rm dma-s2}\rangle=-1.00$ -- within the error bars -- over the whole $\Delta$-range (see panel (B) of Fig.~\ref{fig:domain_sizes}). In contrast, in dmo-s1/s2, just as in dmo-center, the formation of boxes is energetically disfavored and, additionally, the formation of roof-shingle motifs is not possible due to the offset edge contact. Hence, dmo-s1/s2 do not form any coherent tiling. 

A completely different tiling scenario is observed for checkers-sc: 
In checkers-sc, besides a full-edge-contact p-tiling (see cluster 4-p in the corresponding panel of Fig.~\ref{fig:phases}, lower part), off-center patch positioning additionally allows for an offset p-tiling brick configuration (see cluster 4-p-brick in the corresponding panel of Fig.~\ref{fig:phases}, lower part). Subsequently, checkers-sc form a p-tiling with occasional brick-like defects; note that $\langle\Psi_{\rm checkers-sc}\rangle = 1.00$ over the whole $\Delta$-range (see panel (B) of Fig.\ref{fig:domain_sizes}) as np-bonding between two platelets is energetically disfavored (see cluster 2-np in the corresponding panel of Fig.~2, lower part).

\subsubsection{Tilings of asymmetric off-center topologies.}
The visual analysis of our results clearly suggests that the as1 topology does not lead to any tiling (see the corresponding simulation snapshots in Fig.~\ref{fig:phases}, upper part):  both dma-as1 and dmo-as1 allow the formation of fully bonded p- and np-clusters of two and three particles respectively, but these are unable to tile due to overlaps (see the corresponding small cluster analysis in Fig.~\ref{fig:phases}, lower part).

\begin{figure*}
\begin{center}
\includegraphics[width=1.0\textwidth]{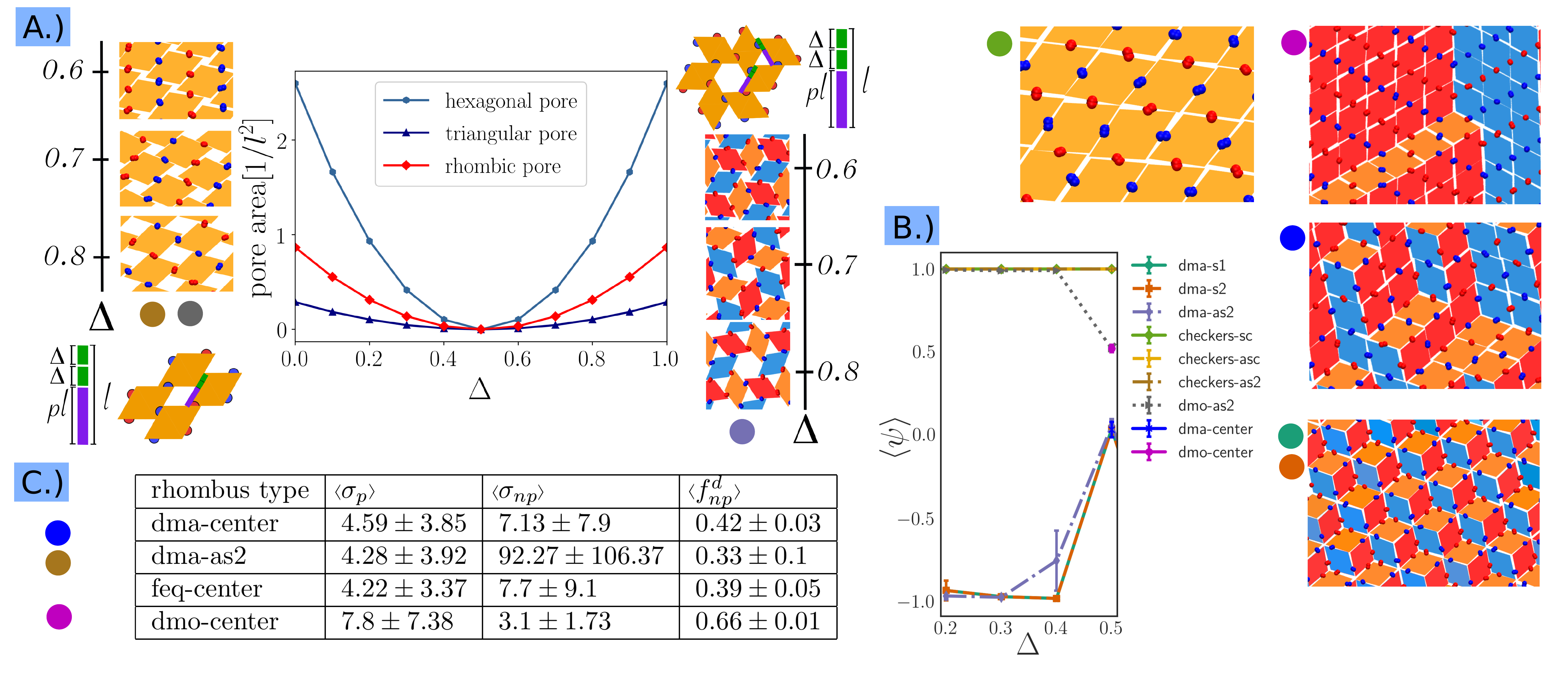}
\caption{Panel (A): Area of rhombic, hexagonal, and triangular pores (symbols and colors as labeled) as function of the patch position $\Delta$ for dmo-as2 (snapshots on the left-hand-side of the graph), checkers-as2 (refer to the dmo-as2 snapshots) and dma-as2 (snapshots on the right-hand-side of the graph). Sketches of pores and their side length $pl$ in the sticky limit are reported on the left- and right-hand-side of the graph, respectively. Panel (B): average order parameter $\langle\Psi\rangle$ as function of $\Delta$, where -1 denotes a completely np-tiling, 1 a completely p-tiling and 0 a random tiling. Equivalent topologies $i$ and $k$ where $\Delta_{i}= 1- \Delta_{k}$ are summed up to one point in order to enhance the statistics. All systems yielding to a tiling are reported, as labeled, and at least one snapshot per tiling is shown, labeled with a large dot, whose color corresponds to the legend of panel (B). Panel (C):  average domain sizes of p- and np-bonded ($\langle \sigma_p \rangle $ and $\langle \sigma_{np}\rangle$) rhombi, and average fraction of np-domains ($\langle f^d_{np}\rangle$) over the whole sample (defined as the number of np-domains over the total number of domains). Note that the domain sizes are distributed exponentially (see Fig.~\ref{fig:domain_size_histo} in Appendix~\ref{domain-sizes}), explaining the extent of the standard deviation. For dma-as2 the domain size is calculated at $\Delta=0.4$ and 0.6.}
\label{fig:domain_sizes}
\end{center} 
\end{figure*}

In contrast to checkers-sc, checkers-asc -- defined by placing patches of one type asymmetrical, while patches of the other type remain in the center --  can only assemble into a brick-tiling  (see clusters 2-p-r and 2-p-b that merge into cluster 4-p-brick in the corresponding panel of Fig.~\ref{fig:phases}, lower part) with $\langle\Psi_{\rm checkers-asc}\rangle = 1.00$  over the whole $\Delta
$-range (see panel (B) of Fig.~\ref{fig:domain_sizes}).

On the other hand, the as2 topology yields open tilings for all three system types (dma, dmo and checkers). As the small cluster analysis shows, while dma-as2 can tile both p- and np-clusters, dmo-as2 and checkers-as2 can tile only p-clusters: boxes of dmo-as2 particles yield unsatisfied bonds or overlaps, while the checkers-as2 can not even form np-bonds (see the clusters in the corresponding panel of Fig.~\ref{fig:phases}, lower part).
As a result, for dmo-as2 and checkers-as2 the competition between op- and onp-tilings is completely suppressed and the op-tiling is selected, resulting in $\langle\Psi_{\rm dmo-as2}\rangle = \langle\Psi_{\rm checkers-as2}\rangle = 1.00$  over the whole $\Delta$-range (see panel (B) of Fig.~\ref{fig:domain_sizes}).

In contrast, for dma-as2 both open tilings are possible. Nonetheless, an open lattice with mixed bonding is never observed for $\Delta\in[0.2, 0.3, 0.7, 0.8]$ since for these $\Delta$-values the connection of p- and np-clusters induces a high strain on the tiling, leading to self-healing processes, $e.g.,$ when a p-cluster forms at the surface of the largest np-cluster during crystallization (see video in the Supplementary Materials). Our results show that the onp-tiling is preferred with respect to the op-tiling, resulting in an order parameter $\langle\Psi_{\rm dma-as2}\rangle=-1.00$  for the aforementioned $\Delta$-values (see panel (B) of Fig.~\ref{fig:domain_sizes}). Nonetheless, the possibility of forming an op-tiling is not completely suppressed: for $\Delta=0.2$, one of eight simulation runs resulted into a op-tiling, while the rest of the simulations yield a onp-tiling. In contrast, for $\Delta\in[0.4,0.6]$ p-defects emerge within the onp-tilings, as indicated by $\langle\Psi_{\rm dma-as2}\rangle=-0.76\pm0.18$ (see panel (B) of Fig.~\ref{fig:domain_sizes}) and the size of the error bar associated to it (see Fig.~\ref{fig:psi_histogram} in Appendix~\ref{energy-impact} ). The presence of p-defects in the np-tiling is also reflected by the finite size of p-domains, where $\langle \sigma_p \rangle \ll \langle \sigma_{np} \rangle$ (see panel (C) of Fig.~\ref{fig:domain_sizes}). Moreover, as soon as $\langle f^d_{np}\rangle < 1$, we know that the tiling is composed of many different domains. 

Within the framework of two patch types, some general trends can be identified. First, the tilings arising from off-center topologies of dma/dmo systems (either symmetric or asymmetric) are very different from their respective center topologies, meaning that on increasing $\Delta$ from 0.2 to 0.5, the order parameter grows from -1 (onp/np-tilings) to almost 0 (random tiling) for all systems with the exception of dmo-as2 where it goes from +1 (op-tiling) to 0; in contrast, all checkers topologies yield parallel tilings (see panel (B) of Fig.~\ref{fig:domain_sizes}). We note that with respect to the feq-systems the jump of the order parameter at $\Delta=0.5$ is much sharper. The second general pattern we found is that symmetric patch topologies lead to closed-packed lattices, whereas a very specific asymmetric patch topology (as2) leads to parallel and non-parallel open lattices with pore sizes dependent on $\Delta$. 

As the order parameter only characterizes the geometry of the tilings, another parameter must be identified to quantitatively describe their porosity. The pore sizes of all described open tilings depend solely on $\Delta$ and can be calculated from the side length of the pores, $pl$ (see Methods and Fig.~\ref{fig:packing} in Appendix~\ref{open-packing}).  The dependence of the pore areas on $\Delta$ in the sticky limit is given in panel (A) of Fig.~\ref{fig:domain_sizes}. In the sticky limit the packing fraction of the onp-tiling is slightly smaller than the packing fraction of the op-tiling, however, the measured packing fractions for finite sized patches are the same within error bars (see Fig.~\ref{fig:packing} in Appendix~\ref{open-packing}). 

Finally, we observe that the nucleation of the open lattices occurs $via$ both particle-by-particle growth and hierarchical crystallization (see video in the Supplementary Materials), whereas in systems leading to closed-packed lattices nucleation proceeds solely via single particle attachment~\cite{Whitelam2013}.

\section{Conclusion}

In summary, we have presented a rationalized picture of how shape and bond anisotropy can be combined to drive the assembly of rhombic platelets toward target tilings with specific properties.  By choosing the identity of the interaction sites and their placement on the rhombi edges, we are able to assemble tilings with identical lattice geometry and tunable porosity from a close-packed arrangement to a highly porous, open lattice.

Despite the simplicity of the model, our findings can be applied to the assembly of small tetracarboxylic acids, such as TPTC and NN4A molecules, on a graphite substrate. We note that possible mismatches between the molecular systems and our coarse-grained description might emerge in some cases due to differences either in the steric constraints or in the bonding energies. In the first case, small overlaps between rhombic units might not correspond to overlaps between molecules, thus slightly enlarging the tiling possibilities of the molecules with respect to our patchy platelets. In the latter case, exploratory simulations have shown that a change in the patch-patch interaction energy can affect the balance between p- and np-bonding and hence the final tiling. While the comparison between molecular systems and patchy platelets is already satisfactory, these two possible sources of mismatch set the challenge for a refined coarse-grained model aimed at a more quantitative investigation of molecular tilings. 
 
In contrast, at the colloidal level, we expect our model to completely describe the assembly of patchy platelets with a regular rhombus shape~\cite{Small_2018}. In the colloidal realm, patches are physical or chemical areas on the colloid surface and they can be differentiated by means of, $e.g.$, surface roughness~\cite{Kegel_2012} or DNA strands~\cite{Pine_2012}, while the placement of the patches might result from the use of colloidal joints~\cite{Kraft_2017}. 

In this zoo of molecular and colloidal building blocks, our work gives design directions for the production of materials with tunable porosity and lattice geometry, paving the way to building new interesting materials. In particular, the ability to fine tune the lattice porosity might make it possible to create lattices that can dynamically and reversibly switch between close-packed and open structures. 

\section*{Aknowledgements}
The authors wish to thank Nuno Maulide, Philipp Marquetand, Boris Maryasin, Miriam Unterlass, Peter van Oostrum, Gy{\"o}rgy Hantal and Yuri Lifanov for fruitful and insightful discussions. EB acknowledges support from the Austrian Science Fund (FWF) under Proj. Nos. V249-N27 and Y-1163-N27. Computation time at the Vienna Scientific Cluster (VSC) is also gratefully acknowledged.

\appendix

\begin{figure*}[t]
    \begin{center}
    \includegraphics[width = 1.0\textwidth]{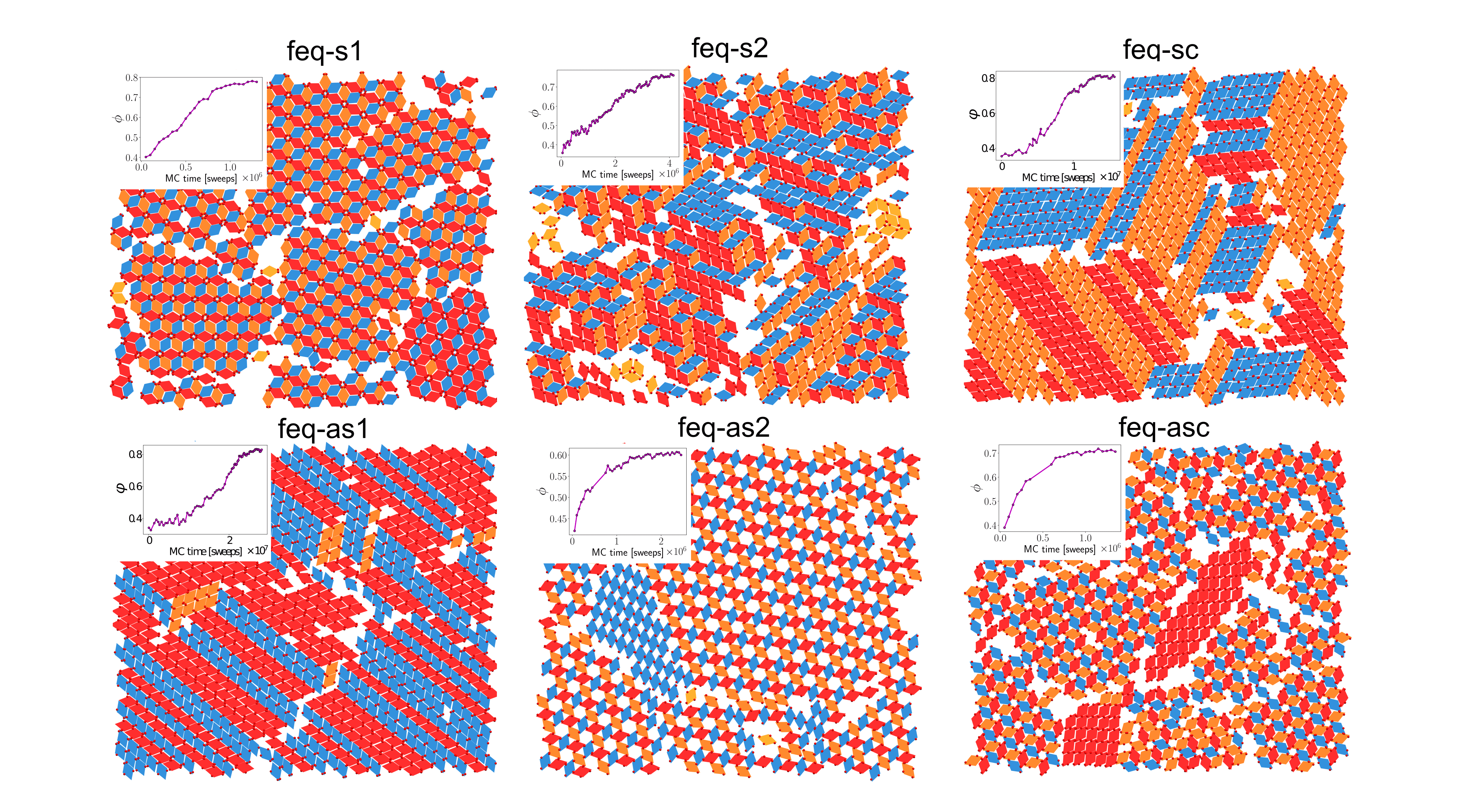}
    \caption{Simulation snapshots of equilibrated feq-systems (main) and packing fraction $\phi$ (inset) at $\Delta=0.2$ and $\epsilon = -5.2k_{B}T$. Particles in the snapshots are colored according to their orientation. However, in some cases colors are mismatched because the particle orientation is calculated with respect to the orientation of the biggest cluster.}
    \label{fig:feq_npt}
\end{center}
\end{figure*}

\section{Patch topologies}
\label{patch-topologies}
\subsection{Topologies for feq systems}
\label{feq-patch-topologies}
We introduce the parameters $\Delta_{k}$ (with $k=1,2,3,$ and 4) that define the distance of patch $k$ from its -- arbitrarily chosen -- reference vertex. If patches can be placed anywhere on the edges, the full set of four parameters $\{\Delta_{k}\}$ is needed to describe one particle type. As we introduce symmetry operations on the patch arrangement, we can describe each particle type by one single parameter $\Delta$ and a patch topology. In general, the investigated topologies can be split into symmetric topologies (s) and asymmetric topologies (as): if we consider the vertex enclosing a $120\degree$ angle as reference vertex for the pair of patches belonging to its respective edges, then in s1 and s2 topologies patch pairs are symmetrically arranged with respect to their shared vertex, while in as1 and as2 topologies pairs of patches are described by $\{\Delta, 1-\Delta\}$. In contrast, in the sc and asc topologies one pair of opposite patches is fixed in the center. We consider three symmetric (s) and three asymmetric (as) patch topologies (see panel (H) of Fig.~\ref{fig:feq_phases}) and $\Delta$-values spaced by 0.1 from $\Delta=0.2$ to 0.8. Note that all six topologies collapse to feq-center for $\Delta = 0.5$ and that particle types are symmetric with respect to $\Delta = 0.5$.

\subsection{Topologies for dma/dmo/checkers systems}
\label{dma-dmo-checkers-patch-topologies}
We consider patchy rhombi with two kinds of patches for any patch arrangement (given again by a patch topology and a $\Delta$-value. Patches of the same kind attract each other with $-\epsilon$ while, patches of a different kind repel each other with $\epsilon$. The three ways to distribute four patches of two types on four rhombus edges are (see the single particle representation in Fig.~\ref{fig:phases}): positioning patches of the same kind to enclose the larger rhombus angles (double-manta, dma), positioning patches of the same kind to enclose the smaller rhombus angles (double-mouse, dmo) and positioning patches of the same kind to sit on the parallel edges (checkers). In all the cases, we define the patch positions as distances to reference vertices. For example, the s1-topology of dma and dmo systems describes all systems where patches of the same kind have the same distance to the vertex enclosed by those patches (see Fig.~\ref{fig:phases}). As in the case of feq-systems, the investigated topologies can be split into symmetric topologies (s), where patches of the same type $m$ retain the same $\Delta$ ($\Delta_{m2} = \Delta_{m1}$) with respect to the reference vertices and asymmetric topologies (as) where $\Delta_{m2} = 1 - \Delta_{m1}$. With these definitions a particle can be characterized, again, by its patch topology and only one $\Delta$. All topologies collapse to their respective central configuration for $\Delta = 0.5$.

In both dma and dmo systems with s2 as well as as1 topology,  $\Delta$ and ($1-\Delta$) particles are equivalent through exchange of patch type, while checkers-sc with $\Delta$ and ($1-\Delta$) are equivalent through rotation (see the corresponding panels in the lower row of Fig.~\ref{fig:phases}). Additionally, even if not denoted in Fig.~\ref{fig:phases}, for dma/dmo/checkers with as2 topology, particles with $\Delta$ and ($1-\Delta$) are equivalent through a three-dimensional flipping. In contrast, for dma-s1, dmo-s1, and checkers-asc systems there is no equivalence operation that turns a $\Delta$ into ($1-\Delta$) particle. 

\begin{figure}[h]
    \begin{center}
    \includegraphics[width = 1.0\columnwidth]{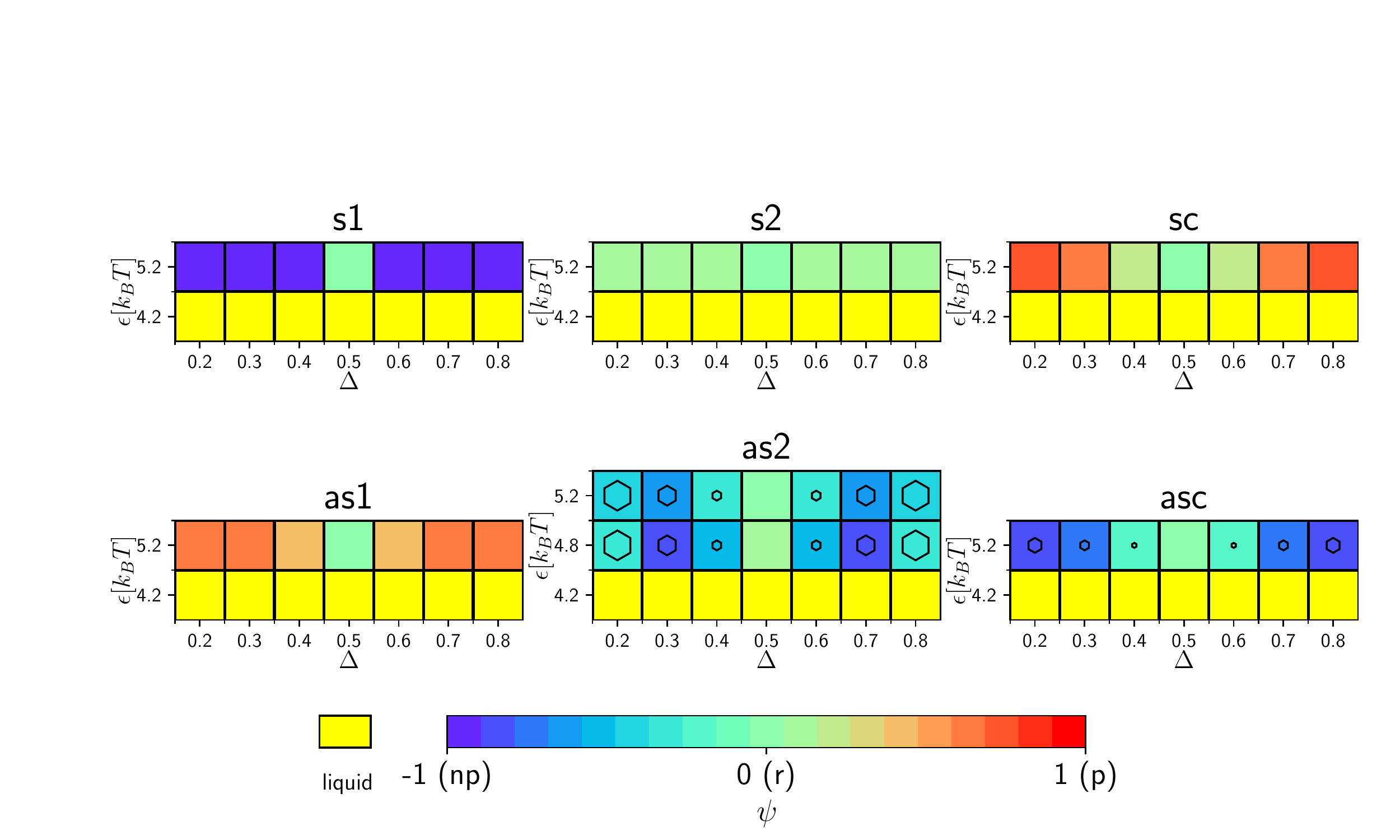}
    \caption{Heatmaps for all feq-systems as function of $\Delta$ and $\epsilon$. Squares are colored according to $\langle\Psi\rangle$. The colors are: yellow (liquid), blue (completely np-tiling), green (random phase), red (completely p-tiling). In the as2 and asc topologies, the sizes of the bold hexagons are proportional to the actual pore sizes in these open systems. Note that phase points for $\Delta > 0.5$ are mirrored at $\Delta = 0.5$ because of particle symmetries.}
    \label{fig:feq_phase_diagram}
  \end{center}
\end{figure}

\begin{figure}[h]
    \begin{center}
    \includegraphics[width = 1.0\columnwidth]{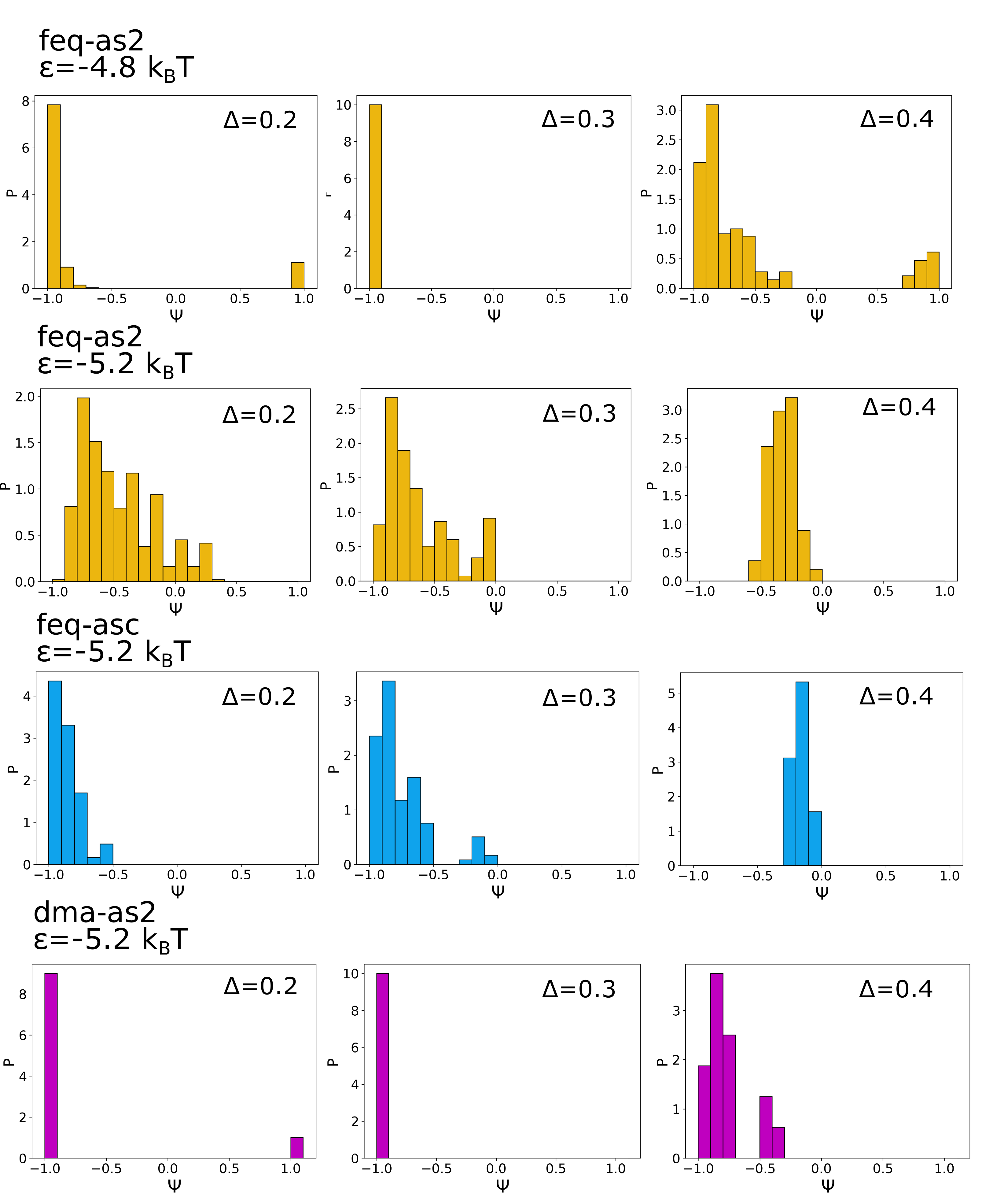}
    \caption{Histogram for order parameter $\Psi$ for the open tilings of feq-as2$_{-4.8k_{B}T}$, feq-as2$_{-5.2k_{B}T}$, feq-asc$_{-5.2k_{B}T}$ and dma-as2$_{-5.2k_{B}T}$ systems (as labeled).}
    \label{fig:psi_histogram}
\end{center}
\end{figure}

\section{Cluster move algorithm}
\label{cluster-move-algorithm}
In a cluster move algorithm with a static linking scheme  \cite{Whitelam2007, Whitelam2010}, a so called pseudocluster is generated by attempting to link particle $i$ with one of its neighbours  $j$ with probability
\begin{equation}
 p_{ij}(\mu\rightarrow \nu) = \Theta(n_{c} -N_{\mathcal{C}})(1-e^{\beta_{f}\epsilon_{ij}}),
\end{equation}
where $\epsilon_{ij}$ is the bonding energy between particle $i$ and $j$, $\beta_{f}$ is a fictitious inverse temperature and $\Theta(n_{c} - N_{\mathcal{C}})$ is a conventional theta function acting as a physically motivated cutoff. A neighbour 
of a particle $i$ is defined as a particle that lies within its interaction range.
Starting from a seed $s$ we try to add all neigbhours of an already linked particle exactly once. A cluster move is
aborted when the pseudocluster size $n_{c}$ exceeds $N_{\mathcal{C}} = \text{rint}(\eta^{-1})$, where $\eta$ is a uniformly distributed random variable between $0$ and $1$. The fictious reciprocal temperature $\beta_{f}$ controls the likelihood of forming 
pseudo-clusters of a certain size. Note that if $\beta_{f}=0$ no pseudo-clusters bigger than one will be formed, whereas if $\beta_{f} = \beta$, the cluster-move acceptance rate is 1. Hence, one can choose $\beta_{f}$ freely between $0$ and $\beta$, and we choose $\beta_{f}=5$.
After generating a pseudo-cluster in the described way, we either translate the particles by the same amount or rotate them around the center of mass of the cluster.
The acceptance criterium for this cluster move is given by
\begin{equation}
W_{acc}(\mu\rightarrow \nu)= \min(1,e^{(\beta_{f} - \beta)[E(\nu)-E(\mu)]}),
\end{equation}
where $E(\nu)$ is the energy of the old state $\nu$ and $E(\mu)$ is the energy of the new state $\mu$.

\section{Order parameters}
\label{order-parameters}
\subsection{Randomness parameter}
\label{randomness-parameter}
The random tiling emerges because both TPTC molecules and feq-center particles are equally likely to bind parallel (p) or non-parallel (np) (see panel (B) of Fig.~\ref{fig:feq_phases}). Nonetheless, as there are binding restrictions on particles attaching to already bonded dimers/multimers, the total number of p- and np-bonds is not equal. To account for this imbalance and to quantify the randomness correctly, the order parameter 
\begin{equation}
\Psi = (0.608 n_{p} - 0.392 n_{np}) / (0.608 n_{p} + 0.392 n_{np})
\end{equation}
was introduced~\cite{Science_2008},  where $n_{p}$ is the total number of p-bonds, $n_{np}$ is the number of np-bonds, while the numerical factors were estimated from simulations~\cite{NatChem_2012}. This order parameter is constructed such that it varies from +1 (p) to -1 (np), where  $\Psi=0$ corresponds to the perfect random tiling estimated in Ref.~\cite{NatChem_2012}.

\subsection{Porosity parameter}
\label{porosity-parameter}
For as2 topologies and in the sticky limit the maximum side length $pl$ of a -- either rhombic or triangular -- pore can be calculated as (see the schematic in panel (A) of Fig.~\ref{fig:domain_sizes})
\begin{equation}\label{eq:pores}
pl = \lvert l - 2\Delta \rvert,
\end{equation}
where $l$ is the length of the rhombus edge. Once $pl$ is known, the areas of the pores can be calculated in a straightforward way for the op- and the onp-tiling (see Fig.~\ref{fig:packing} in Appendix~\ref{open-packing}). 

\begin{figure}[h]
    \begin{center}
    \includegraphics[width = 1.0\columnwidth]{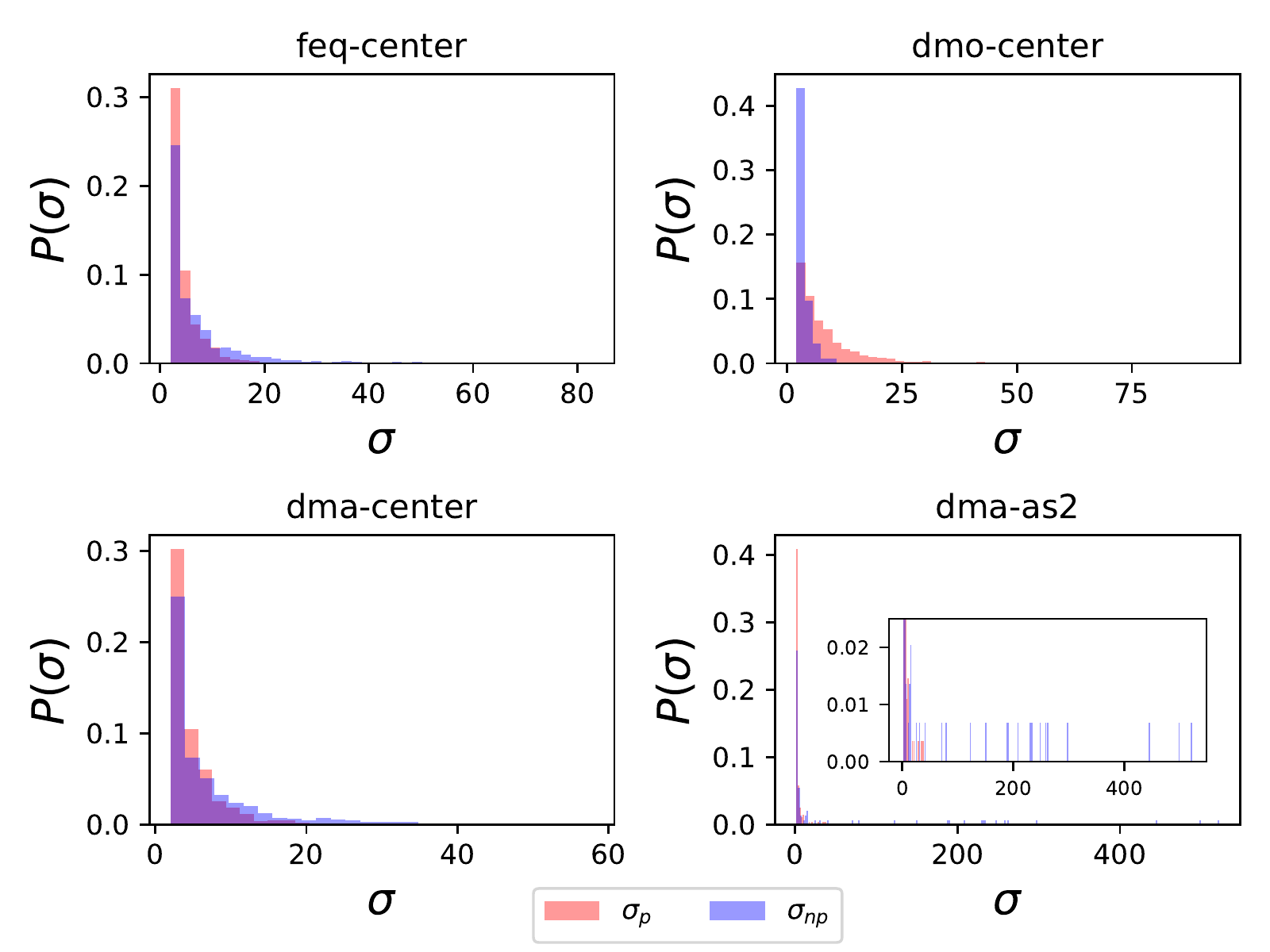}
    \caption{Normalized distributions of parallel (red) and non-parallel (blue) domain sizes for feq ( whitelam comparison), dmo-c, dma-c and dma-as2 at $\Delta$=0.6.}
    \label{fig:domain_size_histo}
 \end{center}
\end{figure}

\section{Equilibration of grand-canonical simulations}
\label{equilibration}
In Fig.~\ref{fig:feq_npt}, we report the evolution of the packing fraction throughout a simulation run for all feq-systems together with a representative snapshot of the resulting tiling. 
We ensure equilibration of all studied systems by calculating the packing fraction $\phi$ over the course of the simulations. All observables, $i.e.$, the order parameter $\langle\Psi\rangle$, the sizes of the parallel $\sigma_p$ and non-parallel $\sigma_{np}$ domains, and the fraction of non-parallel domains $f^d_{np}$ are evaluated after equilibration is reached as averages over the last five simulation checkpoints of the 16 runs per state point.

It is interesting to note that in feq-s1 the frizzy domain boundaries present in the snapshot are due to simultaneous growth of different clusters. In feq-s2 on the other hand, boundaries and holes stem from the system's difficulty to nucleate into a commensurable tiling.  
Although in feq-as2 and, to some degree, in feq-asc multiple clusters grow at the same time, the resulting tilings tend to have few domain boundaries.
These better connected tilings result from the self-healing effect.
In this context it is relevant to note that for feq-as2 and feq-asc we deliberately picked the snapshots with the largest parallel domains, the roughest domain boundaries and (by choosing $\Delta=0.2$) the longest equilibration time.

\begin{figure}[h]
\begin{center}
\includegraphics[width = 1.0\columnwidth]{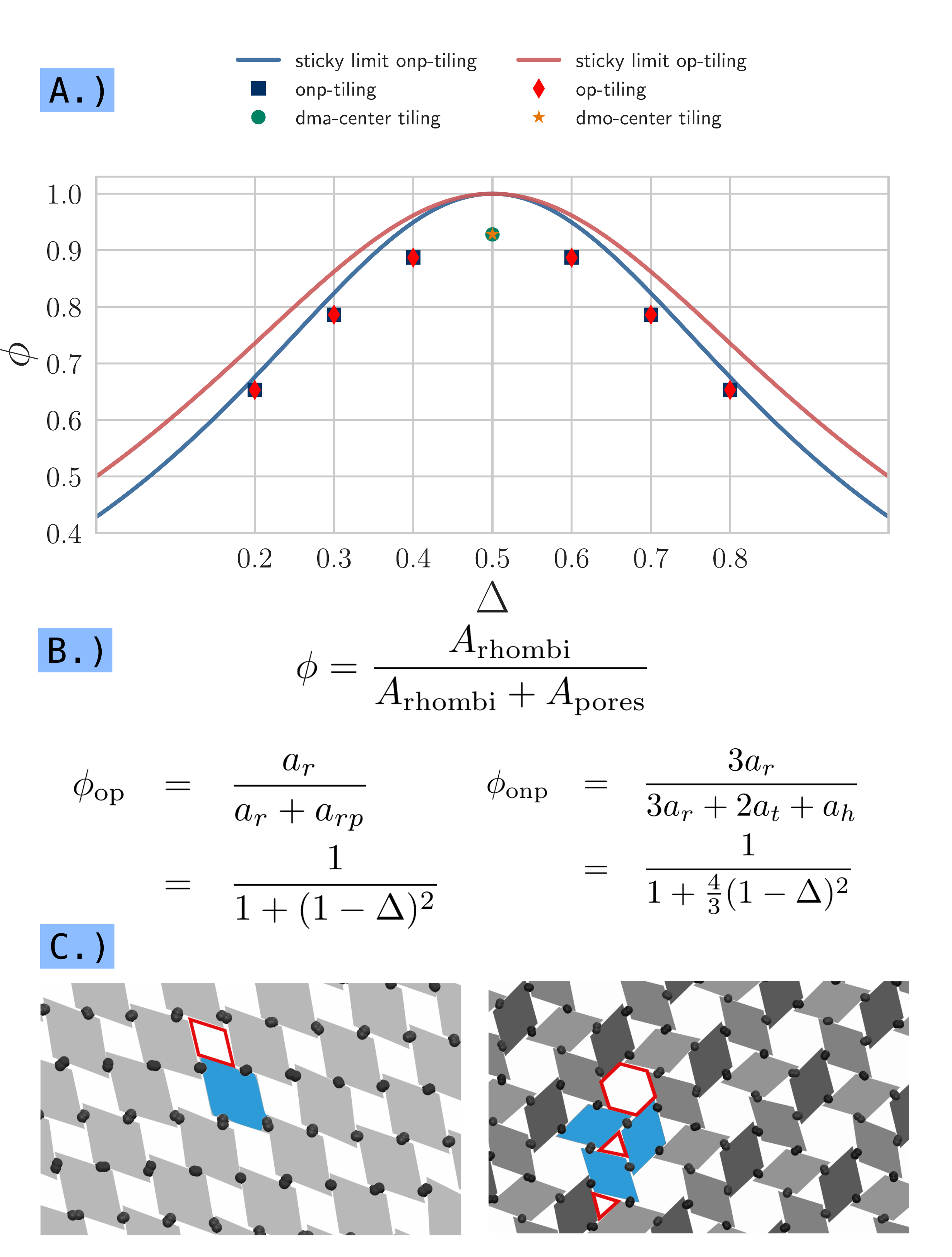} 
\caption{Panel (A): packing fraction $\phi$ as function of $\Delta$, calculated in the sticky limit (solid lines) and measured through pixel counts (points). Panel (B): general formula for $\phi$, where  $A_{\text{rhombi}}$ is the area taken on by particles and $A_{\text{pores}}$ is the area taken on by pores; formula for the packing fraction of the parallel open tiling $\phi_{\rm op}$ in the sticky limit, where $a_{r}$ is the rhombus area and $a_{rp}$ is the area of the rhombic pore; formula for the packing fraction of the open non-parallel tiling $\phi_{\rm onp}$ in the sticky limit, where $a_{r}$ is the rhombus area, $a_{t}$ is the area of the triangular pore and $a_{h}$ is the area of the hexagonal pore. Panel (C): unit cell of the open parallel (left) and open non-parallel (right) tiling.}
\label{fig:packing}
\end{center}
\end{figure}

\section{Patch-patch interaction energy impact on tilings}
\label{energy-impact}
We carry out simulations with different values of the patch-patch attraction energy.

In Fig.~\ref{fig:feq_phase_diagram}, heatmaps show how $\langle \Psi \rangle$ varies depending on topology (s1, s2, sc, as1, as2, asc), the patch position $\Delta$ and the interaction strenght $\epsilon$.
At $\epsilon = -5.2 k_{B}T$ (that is the interaction strength used throughout the paper), feq-systems assemble into tilings with different $\langle \Psi \rangle$. 

In contrast, at $\epsilon = -4.2 k_{B}T$, all feq-systems remained in a liquid phase over a simulation time of $\approx 5\times 10^6$ MC sweeps: at visual inspection of the simulation runs, none of the studied systems formed clusters stable over more than a few MC sweeps and none of the clusters reached sizes bigger than three. 
For feq-as2, we added a state point at $\epsilon = -4.8 k_{B}T$ and it is worth noting that the value of $\langle \Psi_{\rm feq-as2} \rangle$ at $\epsilon = -4.8 k_{B}T$ indicates a more np-tiling for all off-center $\Delta$ than at $\epsilon = -5.2 k_{B}T$.

We note that the sweet spot for the formation of the onp-tiling is confirmed to occur at $\Delta=0.3$, where -- in this case -- p-domains are never observed; in contrast at $\Delta=0.2$ and 0.4, we observe either op-tilings -- seldom --  or -- most of the time --  onp-tilings (see Fig.~\ref{fig:feq_phases} and Fig.~\ref{fig:feq_phase_diagram}). In other words, while at higher bonding energy the growing clusters are mostly a mixture of p- and np-domains, at lower bonding energy p- and np-clusters grow separately and with different probabilities. The interplay between the patch-patch bonding energy and the energy gains/prices of different tilings is very complex and is beyond the scope of the present paper. From our investigations at fixed patch-patch bonding energy, we can conclude that asymmetric patch topologies where patches are relatively far from each other ($i.e.$, for feq-as2 and feq-asc), two different kinds of porous tilings -- characterized by different pore shapes -- emerge and compete, with a clear preference towards a non-parallel pattern.

In Fig.~\ref{fig:psi_histogram} we show the histogram of $\Psi$ for
feq-as2$_{-4.8k_{B}T}$, feq-as2$_{-5.2k_{B}T}$, feq-asc$_{-5.2k_{B}T}$ and dma-as2$_{-5.2k_{B}T}$.
For both dma-as2$_{-5.2k_{B}T}$ and feq-as2$_{-4.8k_{B}T}$ we observe a switch at $\Delta=0.2$: 15/16 runs assemble to a non-parallel open (onp) lattice, while 1/16 runs shows a parallel open (op) lattice. Moving on to $\Delta = 0.3$, 16/16 runs tile non-parallel, indicating a sweet spot for the formation of the non-parallel (np) tiling in both systems. 
At $\Delta=0.4$, both systems become more random, where dma-as2$_{-5.2k_{B}T}$ retains a non-parallel trend, while feq-as2$_{-4.8k_{B}T}$ shows a softened bimodal distribution. 
In contrast, feq-as2$_{-5.2k_{B}T}$ and feq-asc$_{-5.2k_{B}T}$ show an unimodal non-parallel trend for all $\Delta$, where the peaks shift towards a more random $\Psi$ as $\Delta$ increases from 0.2 to 0.4.

\section{Domain sizes of random tilings}
\label{domain-sizes}

In Fig.~\ref{fig:domain_size_histo} we report the size histogram of p- and np-domains for tilings with mixed bonding.
We calculated the domain sizes of p- ($\sigma_{p}$) and np-bonds ($\sigma_{np}$) by interpreting bonds of the same alignment as a network, with particles as vertices and bonds as edges. The p/np-domain sizes are then given by counting the connected components in those networks. Since we count p/np bonds rather than particles, particles can be part of a parallel connected component and a non-parallel connected  component at the same time.
For this reason, the bin height of the histograms (each normalized with respect to the total number of either p- or np-domains) does not give information about the relative amount of p- and np-bonds. It only displays how many more domains of smaller size than bigger size exist within one alignment.
The domain size calculation also yields the fraction of p- ($f^{p}$), and respectively,
the fraction of np-domains ($f^{np}$).

In dma-center, $\sigma_p$, $\sigma_{np}$ and $f^d_{np}$ have values similar to those of the feq-center, reflecting the randomness of the tiling formed by dma-center. In contrast, in dmo-center p-domains are -- on average -- larger than np-domains, whose typical size is $\sigma_{np} \approx 3\pm 2$ particles, due to roof-shingle motives. This is also reflected in the $f^d_{np}$-value that is bigger for dmo-center than for dma-center (see panel (C) of Fig.~\ref{fig:domain_sizes}).

We find that for feq-ceneter, dma-center and dmo-center domain size distributions tend to power-law tails, which is expected for random tilings~\cite{NatChem_2012} and explains the large standard deviation.
In dma-as2 at $\Delta=0.4$ and 0.6, the variance in domain sizes does not stem from random tiling motives (as for feq-center and dma-center) or from roof-shingle motives (as for dmo-center) but from p-defects within the np-tiling. 
In contrast to a defect-free np-tiling, we do observe p-domains of finite sizes: these p-defects can be large enough to split the np-domains in multiple large but disconnected domains.

\section{Packing fraction of open tilings}
\label{open-packing}

In Fig.~\ref{fig:packing}, we show the packing fraction of the open tilings as a function of the patch position $\Delta$ and the unit cells (see panel (A)). The unit cell of the open parallel (op) tiling  consists of one rhombus particle and one rhombic pore, while the unit cell of the open non-parallel (onp) tiling consists of three rhombus particles, two triangular pores and one hexagonal pore (both unit cells are depicted in panel (C)). 
We calculate the packing fraction in the sticky limit (lines in panel (A)) by identifying the unit cells of the open systems and calculating how much space was taken up by particles versus how much space was taken up by pores (see the formulas in panel (B)). 
We measure the packing fraction with finite patch radius ($r_{p} = 0.05$ see Methods section) by taking a sample of gray scale simulation snapshots of the lattices for every $\Delta$, numerically counting the number of non-white pixels (particles) and taking the fraction (see the resulting points in panel (A)). 


\bibliography{references.bib}
\bibliographystyle{unsrt}  


\end{document}